\documentclass[APA,LATO1COL]{WileyNJD-v2}
\usepackage{moreverb}
\usepackage{amsmath}
\usepackage{amsfonts}
\usepackage{amssymb}
\usepackage{graphicx}
\usepackage{caption}
\usepackage{subcaption}
\usepackage{enumerate}
\usepackage{xcolor}
\usepackage{natbib} 
\usepackage{url} 
\usepackage{bm}
\usepackage{hyperref}
\usepackage{cleveref}
\usepackage{amsmath}
\usepackage{multicol}

\usepackage{lipsum}

\hypersetup{breaklinks=true,
            pdfauthor={},
            pdfkeywords = {},
            pdftitle={},
            colorlinks=true,
            citecolor=blue,
            urlcolor=blue,
            linkcolor=magenta,
            pdfborder={0 0 0}}

\newcommand\BibTeX{{\rmfamily B\kern-.05em \textsc{i\kern-.025em b}\kern-.08em
T\kern-.1667em\lower.7ex\hbox{E}\kern-.125emX}}

\articletype{Article}%

\received{<day> <Month>, <year>}
\revised{<day> <Month>, <year>}
\accepted{<day> <Month>, <year>}

\begin{document}

\title{Bayesian Adaptive Polynomial Chaos Expansions}

\author[1]{Kellin N. Rumsey*}
\author[1]{Devin Francom}
\author[1]{Graham Gibson}
\author[2]{J. Derek Tucker}
\author[2]{Gabriel Huerta}

\authormark{Rumsey et al.}

\address[1]{\orgdiv{Statistical Sciences}, \orgname{Los Alamos National Laboratory}, \orgaddress{\state{NM}, \country{United States}}}
\address[2]{\orgdiv{Statistical Sciences}, \orgname{Sandia National Laboratory}, \orgaddress{\state{NM}, \country{United States}}}


\corres{*Kellin Rumsey, \email{knrumsey@lanl.gov}}

\presentaddress{P.O. Box 1663 \\
Los Alamos, NM 87545}

\abstract[Abstract]{
Polynomial chaos expansions (PCE) are widely used for uncertainty quantification (UQ) tasks, particularly in the applied mathematics community. However, PCE has received comparatively less attention in the statistics literature, and fully Bayesian formulations remain rare—especially with implementations in \textsf{R}. Motivated by the success of adaptive Bayesian machine learning models such as BART, BASS, and BPPR, we develop a new fully Bayesian adaptive PCE method with an efficient and accessible \textsf{R} implementation: \texttt{khaos}. Our approach includes a novel proposal distribution that enables data-driven interaction selection, and supports a modified $g$-prior tailored to PCE structure. Through simulation studies and real-world UQ applications, we demonstrate that Bayesian adaptive PCE provides competitive performance for surrogate modeling, global sensitivity analysis, and ordinal regression tasks.
}


\jnlcitation{\cname{%
\author{Rumsey, K. N.}, 
\author{Francom, D.}, 
\author{Gibson, G. C.}, 
\author{Tucker, J. D.}, 
\author{Huerta, G.}} (\cyear{2025}), 
\ctitle{Bayesian Adaptive Polynomial Chaos Expansions}, \cjournal{Stat}, \cvol{2025.aa.bb}.}

\maketitle

\section{Introduction}

Polynomial chaos expansions (PCE), originally described by \citet{wiener1938homogeneous}, have become a widely used tool for surrogate modeling and uncertainty quantification (UQ), particularly in fields such as physics, engineering, and applied mathematics \citep{ghanem1991stochastic, xiu2002wiener, novak2018polynomial}. PCEs represent the response surface of a computer model as a linear combination of tensor products of orthogonal polynomials in the model’s input variables. By projecting model outputs onto these polynomial bases, PCE provides a functional approximation of the input-output relationship. The technique has a long and established history, particularly for propagating uncertainty in simulations involving physical systems \citep{luthen2021sparse}. PCE is also widely used for global sensitivity analysis, where Sobol or derivative-based indices can be derived analytically from the polynomial coefficients \citep{sudret2008global, shao2017bayesian}.

Despite its strengths, the broader use of PCE in statistical modeling has been somewhat limited by concerns related to overfitting in high-degree expansions, challenges with uncertainty quantification, and sensitivity to input distributions \citep{o2013polynomial}. At the same time, recent years have seen the success of fully Bayesian, nonparametric regression tools such as Bayesian additive regression trees (BART; \citealp{chipman2010bart}), Bayesian adaptive spline surfaces (BASS; \citealp{denison1998bayesian, francom2020bass}), and Bayesian projection pursuit regression (BPPR; \citealp{collins2024bayesian}). These models provide flexible, adaptive representations of complex surfaces, while offering natural uncertainty quantification and strong empirical performance across a variety of tasks.

Inspired by these developments, we propose a new fully Bayesian implementation of adaptive PCE. The method builds polynomial basis functions incrementally using a Reversible Jump Markov Chain Monte Carlo (RJMCMC) algorithm. This allows the model to adapt its complexity to the data, enabling a dynamic balance between parsimony and flexibility. A novel proposal distribution governs the selection of interaction terms, leading to efficient exploration of the model space. We also consider a modified $g$-prior for the regression coefficients, which induces shrinkage based on the complexity of a basis function and leverages a Laplace approximation for fast and tuning-free inference.

The rest of this article is organized as follows. \Cref{sec:Background} reviews relevant background on PCE and the sparse Bayesian PCE approach of \cite{shao2017bayesian}. \Cref{sec:adaptive} develops our proposed model, KHAOS (implementation in R at \url{https://github.com/knrumsey/khaos}). A simulation study comparing the method to several popular alternatives is presented in \cref{sec:simulations}, and sensitivity analyses conducted with KHAOS are presented in \cref{sec:application} for two real-world datasets. Concluding remarks are given in \cref{sec:conclusion}.


\section{Polynomial Chaos Expansions}
\label{sec:Background}
\subsection{PCE Framework}
In PCE, a function $f(\bm x)$ with input variables $\bm x \in [0,1]^p$ is approximately represented as
\begin{equation}
\label{eq:pce}
    f(\bm x) \approx \sum_{m=0}^M \beta_m \prod_{j=1}^p \psi_{\alpha_{mj}}(x_j),
\end{equation}
where $\psi_\alpha()$ is the \emph{standardized shifted-Legendre polynomial} of degree $\alpha$. These orthogonal polynomials are equal to $\psi_\alpha(x) = \sqrt{2\alpha+1}P_\alpha(2x-1)$ where $P_\alpha()$ are the Legendre polynomials which satisfy the recurrence relation $(\alpha+1)P_{\alpha+1} = (2\alpha+1)xP_\alpha(x) - \alpha P_{\alpha-1}(x)$ with $P_0(x) = 1$, $P_1(x) = 1$. We note that more general definitions exist, but the above is sufficient for our purposes. 

For a basis function with multi-index $\bm\alpha = (\alpha_1, \ldots, \alpha_p) \in \mathbb N^p$, the \emph{degree} is $d(\bm\alpha) = \sum_{j=1}^p \alpha_j$ and the \emph{order} is $q(\bm\alpha) = \sum_{j=1}^p 1(\alpha_j > 0)$, where $1()$ is the indicator function. A PCE representation is said to be \emph{full} with respect to degree $d$ and order $q$ if all coefficients are non-zero and it contains a term for every multi-index $\bm\alpha$ in the set 
\begin{equation}
\label{eq:A_set}
    \mathcal A_{p,d,q} = \{\bm\alpha \in \mathbb N^p : d(\bm\alpha) \leq d \text{ and } q(\bm\alpha) \leq q\}.
\end{equation}
A PCE is said to be \emph{sparse} if it contains terms for only a subset of $\mathcal A_{p,n}$ (or equivalently, if any of the coefficients are exactly zero). We note that, for PCE models with maximum degree $d$ and maximum order $q$, there are 
\begin{equation}
\label{eq:A_cardinality}
    |\mathcal A_{p,d,q}| = \sum_{i=1}^q\sum_{j=1}^d\binom{p}{i}\binom{j-1}{i-1} = \mathcal O\left(\frac{(pd)^q}{(q!)^2} \right) = \mathcal O\left(\left( \frac{pd}{q^2}\right)^q\right)
\end{equation}
permissible basis functions. 

For this to remain feasible for even moderately sized input dimensions ($p$), one must either (i) place restrictions on $d$ and/or $q$, or (ii) induce a high level of sparsity. A wide range of solvers have been proposed for sparse PCE, including convex optimization methods such as LASSO and LARS, greedy stepwise algorithms like orthogonal matching pursuit, and Bayesian compressive sensing approaches based on variational inference or EM algorithms (see \cite{luthen2021sparse} for an extensive review). Most of these approaches rely on point estimates and cross-validation to select model complexity, and do not provide full posterior uncertainty quantification.

Fully Bayesian approaches to sparse PCE are less common. One recent example is the method of \citet{shao2017bayesian}, which combines a likelihood-based model with sparsity-inducing priors and uses a forward-selection algorithm for model construction. While this approach does not sample from the full posterior distribution, it borrows strength from Bayesian modeling and offers a computationally efficient alternative to traditional MCMC. In the following section, we briefly review this approach, which we include in the simulation study of \cref{sec:simulations}.

\subsection{Sparse Bayesian PCE}
In this section, we briefly describe the algorithm proposed by \cite{shao2017bayesian} (SBPCE) and we discuss a few optional modifications which are available in the \texttt{khaos} implementation. This algorithm is not fully Bayesian in the sense that $M$ and $\bm\Psi$ are determined algorithmically rather than being inferred as part of the posterior. The SBPCE approach proceeds as follows:
\begin{enumerate}
    \item For fixed maximum degree $d_\text{max}$ and maximum order $q_\text{max}$, generate the complete set of $|\mathcal A_{p,d_\text{max},q_\text{max}}|$ basis functions.  
    \item Initialize a model which returns the sample mean $(y_1+\cdots + y_n)/n$ for all $\bm x$. 
    \item For each basis function, compute the sample correlation $r_m = \text{cor}(\bm\psi_m(\bm x | \bm \alpha_m), \bm y)$ and reorder the basis columns so that $r_m^2 \geq r_{m+1}^2$. 
    \item For each basis function, compute the squared partial correlation component $\rho_{m|1,\ldots, m-1}^2$. Reorder the basis functions again so that $\rho_{m|1,\ldots, m-1}^2 \geq \rho_{m+1|1,\ldots, m}^2$.
    \item For every $m \in \{0,\ldots,M\}$, consider the model $\mathcal M_m$ with basis functions $\bm \psi_m, \ldots, \bm\psi_0$. Take $m^\star$ to be the largest M such that the Kashyap information criteria (KIC) for model $\mathcal M_m$ is larger than that of $\mathcal M_{m+1}$. 
    \item {\bf Enrichment:} If $\mathcal M_{m^\star}$ model contains a maximally complex term (i.e. one with degree $d_\text{max}$ and/or order $q_\text{max}$), then we (i) increment $d_\text{max}$ and/or $q_\text{max}$, (ii) enrich the set of candidate basis functions and (iii) return to step 2. Otherwise, return $\mathcal M_{m^\star}$.
\end{enumerate}
The original enrichment scheme of SBPCE is quite restrictive, leading to a fast and parsimonious training algorithm. Unfortunately, it can permanently cut out certain input variables and leads to a strong dependence on the initial choice of $d_\text{max}$ and $q_\text{max}$. In appendix A of the supplement, we discuss several alternative enrichment strategies which can improve the accuracy of the SBPCE approach (and reduce dependence on tuning-parameters) at the cost of increased computation. In \cref{sec:adaptive_gprior}, we also show how step $5)$ can be replaced with a closed form Bayes Factor based on the modified g-prior. 

\subsection{Sobol Indices}
One appealing feature of PCEs, is that they make it easy to compute Sobol indices, which are widely used for global sensitivity analysis \citep{sobol2001global, sudret2008global, francom2018}. 

In a Sobol analysis, the function of interest is assumed to admit an ANOVA-like decomposition:
$$f(\bm x) = f_0 + \sum_{i=1}^pf_i(x_i) + \sum_{i < j}^p f_{ij}(x_i, x_j) + \ldots + f_{1,\ldots, p}(x_1, \ldots, x_p) = \sum_{m=0}^M f_{{\bm u}_m}(\bm x_{\bm u_m}),$$
with every term being orthogonal and centered at zero (except for $f_0$). It follows that the variance of $f(\bm x)$ can then be decomposed as
$$\text{Var}(f(\bm x)) =  \sum_{i=1}^p V_i + \sum_{i < j}^p V_{ij} + \ldots + V_{1,\ldots, p} = \sum_{m=1}^M V_{\bm u_m}.$$
The $V_{\bm u}$ terms are usually rescaled (so that they sum to unity) as $S_{\bm u} = V_{\bm u} / \text{Var}(f(\bm x))$ and called \emph{partial sensitivity indices}. The \emph{total sensitivity index} for the $i^{th}$ input is defined as $T_i = \sum_{\bm u: i \in \bm u} S_{\bm u},$ which are only guaranteed to sum to at least $1$.

The main insight is that, by construction, PCE models are already expressed in this orthogonal form—assuming the inputs are independent and uniformly distributed on \([0,1]\). In particular, each term in the PCE expansion can be associated with a specific subset $\bm u$ of input variables, and the contribution to the variance is
$
V_{\bm u} = \sum_{m \in \mathcal{A}_{\bm u}} \beta_m^2,
$
where $\mathcal{A}_{\bm u}$ indexes all basis functions that depend on exactly the variables in $\bm u$. In words, the partial sensitivity index for a subset $\bm u$ is the sum of squared coefficients for all PCE terms that involve exactly those variables. For further discussion, see \citet{sudret2008global}.

\section{Adaptive Bayesian PCE}
Following the principle of NUAP (no unnecessary acronyms please; \cite{nuap2011}), we avoid labeling our approach with a cumbersome acronym. Instead, we refer to this method as KHAOS, in reference to the \texttt{khaos} R package that implements it, which was named in turn for the primordial void of Greek mythology (\url{https://github.com/knrumsey/khaos}). Despite the name, the KHAOS algorithm (or model, or approach) refers simply to the adaptive Bayesian polynomial chaos expansion described in this section. 

\label{sec:adaptive}
\subsection{The KHAOS Model}

Let $y_i$ denote the response variable and $\bm x_i$ denote a vector of $p$ covariates ($i=1,\ldots, n$). Without loss of generality, we assume that $\bm x \in [0,1]^p$. The response is modeled as
\begin{equation}
    \begin{aligned}
        y_i &= f(\bm x_i) + \epsilon_i, \quad \epsilon_i \sim N(0, \sigma^2) \\
        f(\bm x) &= \beta_0 + \sum_{m=1}^M\beta_m\Psi_m(\bm x|\bm\alpha_m) \\
        \Psi_m(\bm x|\bm\alpha_m) &= \prod_{i=1}^p\psi_{\alpha_{mj}}(x_j),
    \end{aligned}
\end{equation}
where each Basis function $\Psi_m$ is fully defined by the multi-index $\bm\alpha_m$ (described in \cref{sec:Background}). We define $\bm A = \{\bm \alpha_1, \ldots, \bm\alpha_M\}$ and specify the prior for the basis function parameters $(\bm\alpha_1, \ldots, \bm\alpha_M, M)$ as
\begin{equation}
\begin{aligned}
    \bm \alpha_m |M & \stackrel{\text{iid}}{\sim} \text{Unif}\left( \mathcal A_{p,d_\text{max}, q_\text{max}} \right),\quad m=1,\ldots, M \\
    M|\lambda &\sim \text{Poiss}(\lambda) \\
    \lambda &\sim \text{Gamma}(a_M, b_M).
\end{aligned}
\end{equation}
Although a prior that penalizes complexity in the multi-indices (e.g., by degree or order) could be specified, we adopt a uniform prior over admissible basis functions and instead encourage parsimony through the modified $g$-prior on the coefficients, as described in \cref{sec:adaptive_gprior}.

For the remaining parameters $(\bm \beta, \sigma^2)$, we specify the prior
\begin{equation}
\label{eq:betasigma}
    \begin{aligned}
        \bm\beta | M, \sigma^2, \bm S_0 &\sim \mathcal N_{M+1}\left(\bm 0, \sigma^2\bm S_0\right) \\
\sigma^2 &\sim \text{Inv-Ga}(a_\sigma, b_\sigma).
    \end{aligned}
\end{equation}
where $\bm S_0$ is a prior covariance matrix whose structure we discuss in \cref{sec:adaptive_gprior}. 

\subsection{Efficient Posterior Sampling}
\label{sec:adaptive_proposal}

Fully Bayesian inference is complicated here by the fact that $M$, the number of basis functions, is allowed to grow and shrink. This requires transdimensional proposals, which we handle using a reversible jump Markov chain Monte Carlo (RJMCMC) algorithm. This framework has seen success in several modern contexts including \citep{francom2020bass, rumsey2024generalized, collins2024bayesian}

At each iteration of the MCMC sampler, we propose to modify the current model using one of four possible moves:
\begin{enumerate}
    \item \emph{Birth}: Propose adding a new basis function.
    \item \emph{Death}: Propose removing an existing basis function.
    \item \emph{Mutation (degree)}: Modify the degree partition of an existing basis function.
    \item \emph{Mutation (variable)}: Swap a variable within an existing basis function.
\end{enumerate}

These moves allow the model to flexibly explore the space of basis configurations. The remaining parameters $(\bm\beta, \sigma^2)$ are updated via Gibbs steps, using their conditional posteriors described in \cref{sec:adaptive_gprior}.

Each proposed move is accepted with probability
\begin{equation}
\log \alpha_X = \log \left( \frac{p(\bm y \mid \mathcal{M}_{\text{cand}})}{p(\bm y \mid \mathcal{M}_{\text{curr}})} \right)
+ \log \left( \frac{p(\mathcal{M}_{\text{cand}})}{p(\mathcal{M}_{\text{curr}})} \right)
+ \log A_X,
\end{equation}
where $\mathcal{M}_{\text{curr}}$ and $\mathcal{M}_{\text{cand}}$ refer to the current and proposed model, respectively. The final term $\log A_X$ accounts for the proposal probabilities specific to move type $X \in \{\text{Birth, Death, Mutate1, Mutate2}\}$. The first two terms correspond to the log-likelihood ratio and the log-prior ratio, respectively. Explicit equations for $p(\bm y|\mathcal M)$ and $p(\mathcal M)$ are given in Appendix B of the supplement. 

For each of the move types (discussed below), the prior ratio simplifies considerably since the difference in $M$ is at most one:
\begin{equation}
    \frac{p(\mathcal{M}_{\text{cand}})}{p(\mathcal{M}_{\text{curr}})} =
    \begin{cases}
        (M+a_M)\left[(M+1)(b_M+1)|\mathcal A_{p,d_\text{max},q_\text{max}}|\right]^{-1}, & \text{Birth} \\
        (M-1+a_M)^{-1}M(b_M+1)|\mathcal A_{p,d_\text{max},q_\text{max}}|, & \text{Death}  \\
        1, & \text{Mutate1, Mutate2}
    \end{cases}
\end{equation}

\subsubsection{Birth Step}
During a birth step, selected with probability $P_B$, we need only propose a new vector of degrees $\bm\alpha^\star$, in order to completely define the new basis function. \cite{nott2005efficient} suggest an efficient proposal that favors choosing variables which are already in the model -- important when $p$ is large and exploring all $2^p$ interactions is not possible. However, their approach requires evaluating Wallenius' non-central hypergeometric distribution, which rapidly becomes computationally burdensome or numerically unstable in many practical settings. As a result, \cite{nott2005efficient} restrict their algorithm to pairwise interactions, while \cite{francom2018} extend it to three way-interactions. We introduce a related approach that achieves similar variable-selection goals without these limitations. Specifically, we use a weighted coin-flipping procedure that avoids the need for Wallenius’ distribution and does not impose a hard cap on the maximum interaction order.

We begin by sampling an expected interaction order $q_0$ from the set $\{1, \ldots, q_\text{max}\}$ with weights proportional to $q_0^{-s_q}$ (default $s_q=1$). Next, we construct the probability $\eta_j$ that $x_j$ will be active in the proposed basis function, such that $\sum_{j=1}^p\eta_j = q_0 = E(q)$. The idea is that $\eta_j \geq \eta_{j'}$ if $x_j$ is more active than $x_{j'}$ in the current model (see appendix C of the supplemental for details). 

We then independently flip a coin for the inclusion of each input, $\chi_j \sim \text{Bern}(\eta_j)$, which gives us the proposed interaction order as $q(\bm\alpha^\star) = \sum_{j=1}^M \chi_j$. The total degree is sampled from the set $\{q(\bm\alpha^\star), \ldots, p\}$ with sampling weights $d^{-s_d}$ (default $s_d=1$), and is randomly partitioned across the $q$ active variables (i.e. those with $\chi_j=1$). This is done so that each suitable partitioning is equally likely, with probability $\binom{d(\bm\alpha^\star)-1}{d(\bm\alpha^\star)-1}^{-1}$.

For the Metropolis-Hastings acceptance ratio, the proposal term can be written as 
\begin{equation}
    A_\text{Birth} = \frac{P_D \ \binom{d-1}{q-1} \ \prod_{j=1}^p\eta_j^{\chi_j}(1-\eta_j)^{1-\chi_j}}{P_B \ (M+1) \ q_0^{s_q} c_q d^{s_d} c_d},
\end{equation}
where $c_q = \sum_{q_0=1}^{q_\text{max}}q_0^{-s_q}$ and $c_d = \sum_{d=q}^p d^{-
s_d}$. Delayed rejection steps are also included to improve efficiency \citep{green2001delayed}; see Appendix C in the supplement for more details.

\subsubsection{Death Step}
During a death step, selected with probability $P_D$, a basis function is randomly selected for deletion. Because this move reduces the model dimension, the reverse proposal corresponds to a birth step — where a specific multi-index $\bm\alpha^\star$ would have been proposed using the weighted coin-flipping strategy described previously. The reverse move’s proposal probability must marginalize over all values of the expected interaction order $q_0$ that could have generated the deleted basis function. 

The full proposal ratio term for the Metropolis–Hastings acceptance probability is then:
\begin{equation}
A_\text{Death} = \frac{P_B \ d^{s_d} c_d}{P_D \ M \ \binom{d-1}{q-1}} \cdot \left[ \frac{1}{c_q} \sum_{q_0=1}^{q_{\max}} q_0^{-s_q} \prod_{j=1}^p \eta_j(q_0)^{\chi_j}(1-\eta_j(q_0))^{1 - \chi_j} \right],
\end{equation}
$q$ and $d$ refer to the interaction order and total degree of the deleted basis function, and $\chi_j$ indicates whether variable $x_j$ was included in that term. 

To account for delayed rejection in the Birth step, we must condition on the fact that certain proposals would been rejected (e.g., those yielding $q=0$ or $q>q_\text{max})$. This requires evaluation of Poisson-Binomial densities (or an efficient normal approximation). See the supplement for additional details.

\subsubsection{Mutate Steps}
When a mutation step is selected (with probability $P_M = 1 - P_B - P_D$), a single basis function is modified without changing the model dimension. Two types of mutation are used:
(i) resampling the degree partition across the active variables, or
(ii) swapping one active variable for a previously inactive one. The probability of selecting each type is adapted throughout the MCMC, based on their empirical acceptance rates, but is never allowed to drop below 10\% for either type (unless $p\leq 3$ in which case variable mutation is unnecessary).

In a degree mutation, we change only the total degree $d$ and randomly repartition it across the $q$ active variables. The acceptance ratio includes the change in proposal density due to the total degree and its partitioning:
\begin{equation}
A_\text{Mutate1} = \frac{d_\text{curr}^{s_d} \cdot \binom{d_\text{curr}-1}{q - 1}}{d_\text{cand}^{s_d} \cdot \binom{d_\text{cand}-1}{q - 1}},
\end{equation}
where $q$ is the (fixed) interaction order, $d_\text{cand}$ is the proposed degree, and $d_\text{curr}$ is the current degree. The two binomial terms reflect the uniform partitioning over the $q$ active variables.

In a variable-swap mutation, one active variable in a basis function is randomly replaced by an inactive one. The proposal distribution is an adaptive categorical distribution, proportional to the current variable inclusion frequencies (plus a fixed baseline). To ensure detailed balance, we compute the Metropolis–Hastings proposal ratio using the forward and reverse selection probabilities:
\begin{equation}
A_\text{Mutate2} = \frac{\tilde{\pi}_{\text{rev}}(x_{\text{old}})}{\tilde{\pi}_{\text{fwd}}(x_{\text{new}})},
\end{equation}
where $\tilde{\pi}_{\text{fwd}}$ and $\tilde{\pi}_{\text{rev}}$ are the normalized empirical inclusion probabilities used to propose the new and old variables, respectively.

\subsection{Gibbs Steps}

Given the current set of basis functions, the remaining model parameters $(\bm \beta, \sigma^2, \lambda)$ can be updated using standard conjugate Gibbs steps. The update for $\lambda$ is
\begin{equation}
    \lambda|\cdot \sim \text{Gamma}(a_M + M, b_M + 1).
\end{equation}
The full conditional posteriors for $\bm \beta$ and $\sigma^2$ are conjugate under all of the priors considered in this work (discussed in the next section). Given the current design matrix $\bm\Psi$, define:
$$\bm\Sigma_n = \left(\bm\Psi^\intercal \bm\Psi + \bm S_0^{-1}\right)^{-1}, \quad \bm\mu_n = \bm\Sigma_n\bm\Psi^\intercal \bm y.$$
Then the Gibbs updates are:
\begin{align}
\bm\beta \mid \cdot &\sim \mathcal{N}\left( \bm\mu_n, \sigma^2 \bm\Sigma_n \right) \\
\sigma^2 \mid \cdot &\sim \text{Inv-Gamma} \left( a_\sigma + \frac{n}{2},
b_\sigma + \frac{1}{2} \sum_{i=1}^n (y_i - \hat y_i)^2 \right),
\end{align}
where $\hat{\bm y} = \bm \Psi \bm\beta$. The prior matrix $\bm S_0$ depends on the choice of coefficient prior. Full specifications for $\bm S_0$ under the ridge prior, $g$-prior, and modified $g$-prior are provided in \cref{sec:adaptive_gprior}.

\subsection{Prior Structure on Coefficients}
\label{sec:adaptive_gprior}
In this section, we describe the prior placed on the regression coefficients, focusing primarily on a modified $g$-prior that allows different levels of shrinkage for different basis terms. The traditional $g$-prior was introduced by \cite{zellner1986assessing} as a computational convenient prior that helps to regularize the coefficients and perform model selection. The $g$-prior is akin to placeing a constant prior on the mean of $\bm y$, rather than on $\bm\beta$ \citep{robert2007bayesian}. Our proposed modification is a ``$n$-component" $g$-prior in the terminology of \cite{zhang2016two}, and seeks to induce stronger regularization on the coefficients for higher-complexity basis functions. 

We begin by defining the vector $\bm g$ with elements 
\begin{equation}
    g_m = \left(\frac{1}{1 + q(\bm \alpha_m)\left[d(\bm\alpha_m) + q(\bm\alpha_m) - 2\right]} \right)^{\zeta/2},
\end{equation}
where $\zeta \geq 0$ is a tuning parameter (with default $\zeta=1$) that controls how strong the penalty for complexity should be. By setting $\zeta = 0$, this method collapses to the traditional Zellner-Siow $g$-prior. Our modified prior is given by
\begin{equation}
\begin{aligned}
        \bm\beta |M, \sigma^2, g_0^2 &\sim \mathcal N_{M+1}\left(\bm 0, \sigma^2g_0^2 \bm D(\bm g)\left(\bm\Psi^\intercal \bm\Psi\right)^{-1}\bm D(\bm g)\right) \\
        g_0^2 &\sim \text{Inv-Gamma}(a_g, b_g), 
\end{aligned}
\end{equation}
where $\bm D(\bm g)$ is the diagonal matrix with the elements of $\bm g$ on its diagonal. This is consistent with \cref{eq:betasigma} with $\bm S_0 = g_0^2 \bm D(\bm g)\left(\bm\Psi^\intercal \bm\Psi\right)^{-1}\bm D(\bm g)$. Although $\bm\Sigma_n$ can be computed directly  in terms of $\bm S_0$, we usually prefer to compute via the 
$$\begin{aligned}
\bm\Sigma_n &= \left(\bm G \odot \bm\Psi^\intercal \bm \Psi\right)^{-1}  \\
\end{aligned}$$
where $\bm G$ is a matrix with elements
$$
\bm G_{m\ell} = \frac{g_0^2g_mg_\ell + 1}{g_0^2g_mg_\ell},$$
which makes obvious the connection to the traditional Zellner-Siow Cauchy $g$-prior, when $\bm g = \bm 1$ \citep{liang2008mixtures}.

The posterior update for the global regularizer $g_0^2$ is based on the conditional posterior
\begin{equation}
\label{eq:gpost}
    \pi(g_0^2|\bm y) \propto g_0^{-2(a_g + M/2)}\text{exp}\left(-b_g/g_0^2\right)\lvert\bm\Sigma_n\rvert^{1/2}.
\end{equation}
There is no easy way to directly sample from \cref{eq:gpost} (unless $\bm g \propto \bm 1$), but an efficient Laplace approximation can be computed based on the inverse gamma distribution (especially when $\bm\Psi^T\bm\Psi \approx n{\bf I}$, which occurs for PCE when the input design is orthogonal). We recommend sampling $g_0^2$ using Metropolis-Hastings, with the Laplace approximation as the proposal distribution. Specifically, we find $(\hat a_g, \hat b_g)$ so that $g_0^2|\bm y \stackrel{\text{aprx}}{\sim} \text{Inv-Gamma}(\hat a_g, \hat b_g)$; using this inverse gamma distribution for the proposal, the acceptance probability becomes
\begin{equation}
\text{min}\left(1, \frac{\pi(g_{0,\text{cand}}^2|\bm y)}{\pi(g_{0,\text{curr}}^2|\bm y)} \frac{\text{IG}(g_{0,\text{curr}}^2|\hat a_g, \hat b_g)}{\text{IG}(g_{0,\text{cand}}^2|\hat a_g, \hat b_g)} \right),
\end{equation}
where $IG(\cdot | a, b)$ denotes the inverse-gamma density with shape $a$ and rate $b$. To see how this prior can be used for the Sparse PCE approach of \cite{shao2017bayesian} (replacing KIC with Bayes Factors based on the modified $g$-prior), see Appendix B of the supplemental materials.

Note that \texttt{khaos} also supports a ridge penalty, i.e. $\bm S_0 = \tau^{-2} {\bf I}$ with $\tau^2$ fixed (default $\tau^2 = 10^5$), which often works quite well for deterministic simulators, but sometimes overfits (or needs tuning) for noisy data. 

\subsection{Laplace Approximations}
\label{sec:adaptive_laplace}
While directly sampling $g_0^2$ from its conditional posterior is challenging, a Laplace approximation provides a fast and robust solution in this setting.  Our strategy will be to construct the approximation under the simplifying assumption that the design matrix satisfies $\bm\Psi^\intercal \bm\Psi = n {\bf I}$, which holds exactly for orthogonal designs on $\bm x$. In many cases, this approximation may be sufficient (especially when using it as a proposal for Metropolis-Hastings). In cases where the orthogonality assumption may not be appropriate, we can instead construct a Laplace approximation to the exact conditional posterior via Newton-Raphson iterations, using the orthogonal solution as an efficient starting place.

Under this simplifying assumption, the conditonal posterior simplifies to 
\begin{equation}
    \label{eq:gpost_orth}
    \pi(g_0^2|\bm y, \text{orthogonal design}) \propto g_0^{-2(a_g + M/2)}\exp{(-b_g/g_0^2)} \prod_{m=1}^M \left(\frac{g_0^2 g_m^2}{1 + g_0^2 g_m^2}\right)^{1/2}.
\end{equation}
The mode of the Laplace approximation can be obtained via fixed-point iteration on a monotonic function $h(g_0^2)$. We start by initializing $\theta_1^\star = b_g/a_g$ and we alternate between computing $G_k$ and $\theta_{k+1}^\star$ where
$$\theta_k^\star = \frac{-a_g + \sqrt{a_g^2 + 4b_gG_k}}{2G_k} \quad G_k = \frac{1}{2}\sum_{m=1}^M \frac{g_m^2}{1 + \theta^\star_{k-1}g_m^2}.$$
We find that this sequence converges rapidly in practice to the mode $m_\theta$. The spread of the approximation is found the usual way: $$s_\theta^2 = \left(-\frac{\partial^2}{\partial\theta^2}\log\pi(\theta|\bm y,\text{orth})\rvert_{\theta=m_\theta}\right)^{-1}.$$
Finally, we solve for the corresponding Inverse Gamma parameters as $\hat a_g = 2 + m_\theta^2/s_\theta^2$ and $\hat b_g = m_\theta\hat a_g$. We find that, especially for computer experiments where Latin hypercube designs are common \citep{mckay1979}, this approximation is sufficient to get good acceptance from Metropolis-Hastings. If needed, however, the more general case can be found using Jacobi's formula and Newton-Raphson iteration. See Appendix D of the supplement for additional details and derivations.

\begin{figure}[ht]
  \centering
  \begin{subfigure}[t]{0.48\textwidth}
    \centering
    \includegraphics[width=\linewidth]{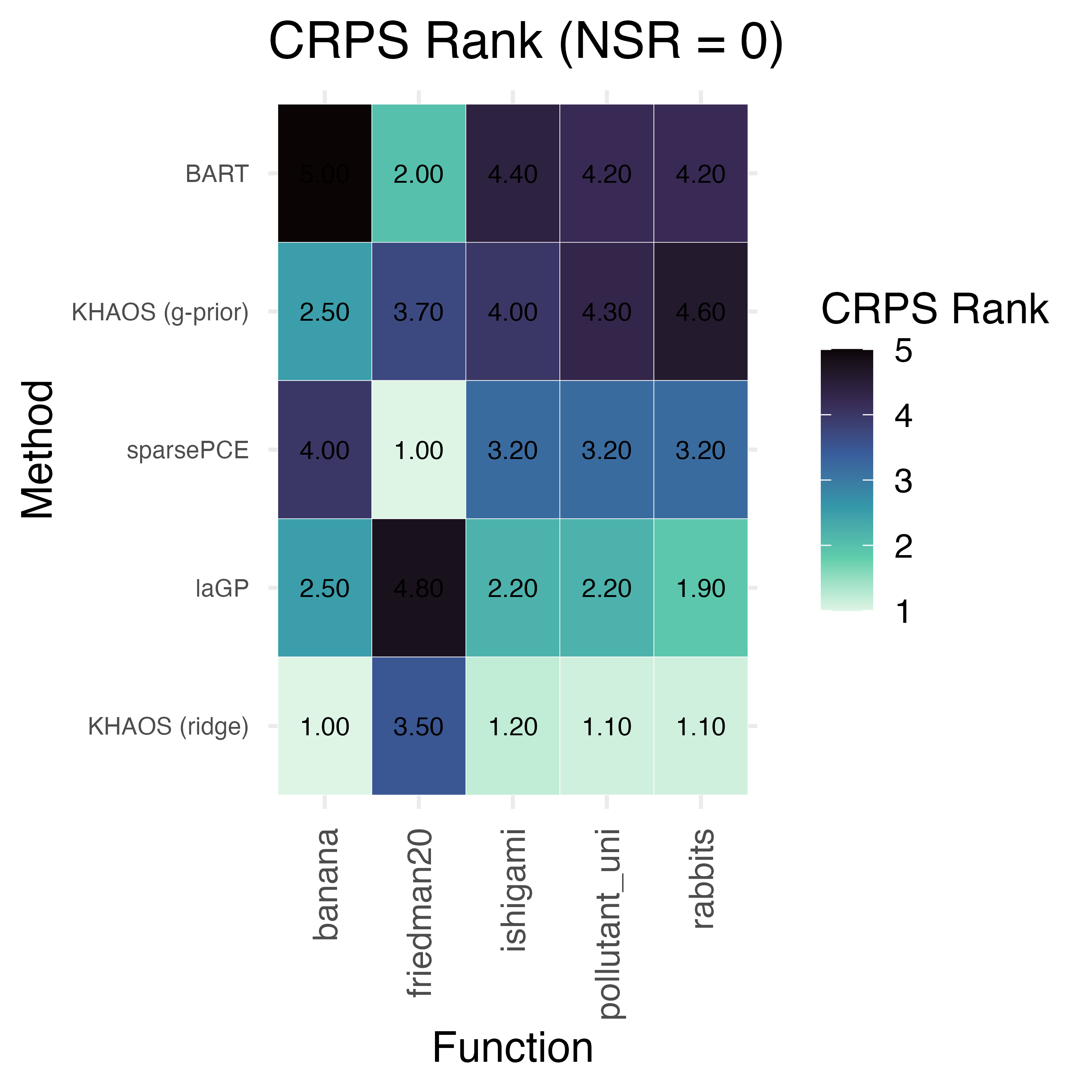}
    \caption{Average CRPS rankings across $10$ replications of each test function. In the noise free ($NSR=0$) setting, KHAOS with a ridge prior has the best average ranking.}
    \label{fig:sub1}
  \end{subfigure}
  \hfill
  \begin{subfigure}[t]{0.48\textwidth}
    \centering
    \includegraphics[width=\linewidth]{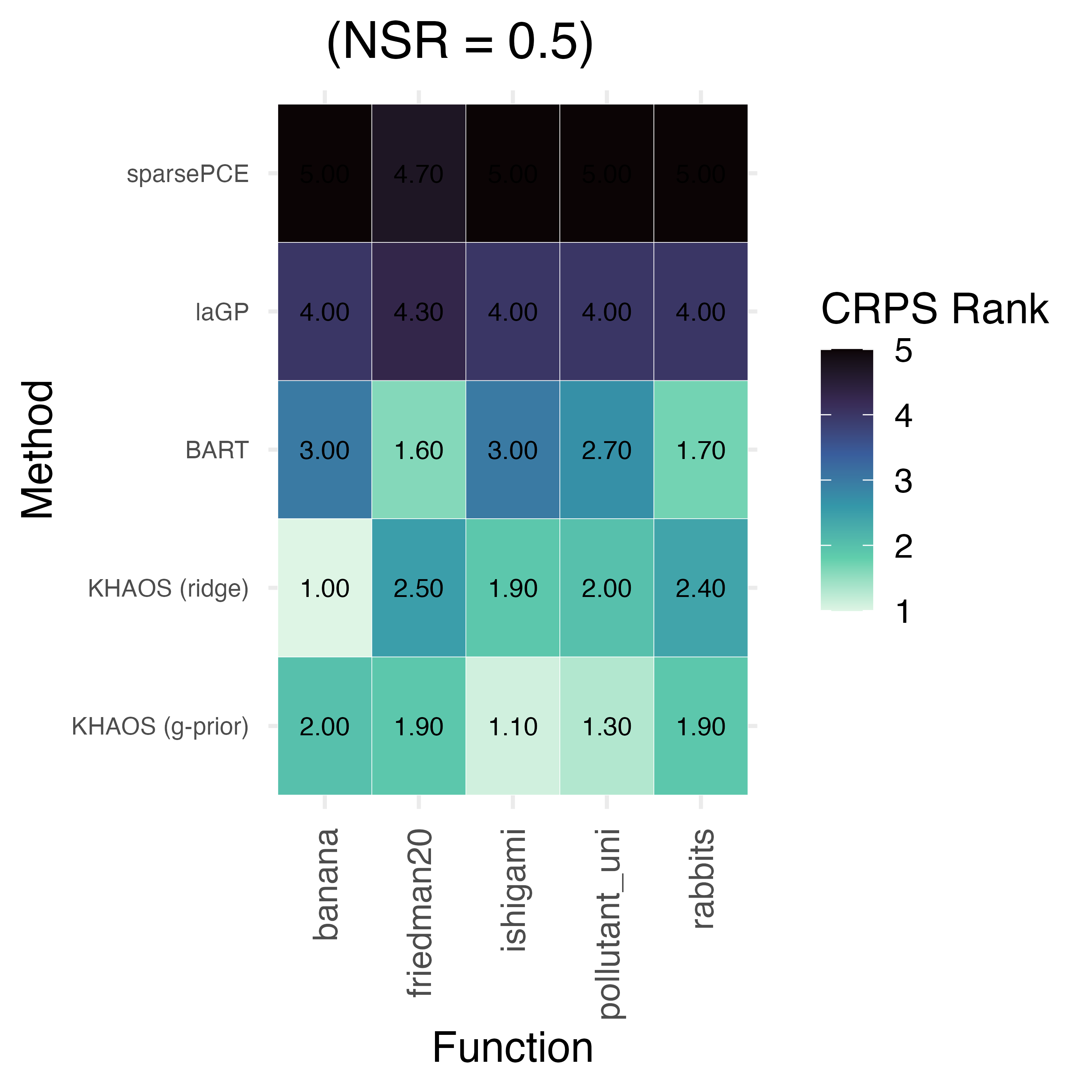}
    \caption{Average CRPS rankings across $10$ replications of each test function. In the high noise ($NSR=0.5$) setting, KHAOS with the modified $g$-prior has the best average ranking.}
    \label{fig:sub2}
  \end{subfigure}
  \caption{}
  \label{fig:combined}
\end{figure}

\section{Simulation Study}
\label{sec:simulations}

We compare the performance of KHAOS under (i) a ridge prior and (ii) the modified \( g \)-prior from \cref{sec:adaptive_gprior}, against several fast competitors. Specifically, we compare to Bayesian additive regression trees (BART; \citealp{chipman2010bart}), the local approximate Gaussian process (laGP; \citealp{gramacy2015}), and a sparse polynomial chaos expansion (PCE) method \citep{shao2017bayesian}, implemented as \texttt{sparse\_khaos} in the accompanying \texttt{khaos} package. This implementation uses a full rebuild enrichment strategy with early stopping to bound the computational complexity. All emulators are run at default settings, and R code for reproduction is included in the supplemental materials.

Simulations are conducted using the \texttt{duqling} R package, designed for transparent and reproducible benchmarking \citep{rumsey2023duqling}. We evaluate the five methods on five test functions:
\begin{itemize}
    \item \textit{banana}: A $p=2$ version of Rosenbrock’s classic banana function.
    \item \textit{ishigami}: A $p=3$ test function commonly used in the PCE literature \citep{ishigami1990importance}.
    \item \textit{rabbits}: A $p=3$ logistic growth model \citep{gotelli2004primer}.
    \item \textit{pollutant\_uni}: A $p=4$ scalar-output model of pollutant diffusion in a river \citep{bliznyuk2008bayesian}.
    \item \textit{friedman20}: A $p=20$ function with only the first five variables active \citep{friedman1991}.
\end{itemize}
See the above references or \texttt{duqling} documentation for further details.

\begin{table}[!b] \centering 
  \caption{Results for the simulation study in the noise free $(NSR = 0)$ setting. The "Within $1\%$ Rate" column gives the proportion of the time that the CRPS of an emulator was within $1\%$ of the best CRPS across all five emulators.} 
  \label{tab:tab2} 
\begin{tabular}{@{\extracolsep{5pt}} lcccc} 
\\[-1.8ex]\hline 
\hline \\[-1.8ex] 
Method & Function & avg. CRPS & avg. Time & Within 1\% Rate \\ 
\hline \\[-1.8ex] 
KHAOS (ridge) & banana & $<0.0001$ & $7.263$ & $1$ \\ 
KHAOS (g-prior) & banana & $1.204$ & $8.855$ & $0$ \\ 
sparsePCE & banana & $6.837$ & $0.049$ & $0$ \\ 
BART & banana & $12.572$ & $7.578$ & $0$ \\ 
laGP & banana & $1.143$ & $20.546$ & $0$ \\[1.2ex] 
KHAOS (ridge) & ishigami & $0.012$ & $13.623$ & $0.9$ \\ 
KHAOS (g-prior) & ishigami & $0.181$ & $11.258$ & $0$ \\ 
sparsePCE & ishigami & $0.066$ & $0.486$ & $0$ \\ 
BART & ishigami & $0.173$ & $7.056$ & $0$ \\ 
laGP & ishigami & $0.030$ & $20.223$ & $0.1$ \\[1.2ex] 
KHAOS (ridge) & rabbits & $0.001$ & $44.251$ & $0.9$ \\ 
KHAOS (g-prior) & rabbits & $0.016$ & $17.578$ & $0$ \\ 
sparsePCE & rabbits & $0.004$ & $0.367$ & $0$ \\ 
BART & rabbits & $0.007$ & $7.869$ & $0$ \\ 
laGP & rabbits & $0.001$ & $20.463$ & $0.1$ \\[1.2ex]
KHAOS (ridge) & pollutant\_uni & $0.0003$ & $12.339$ & $0.9$ \\ 
KHAOS (g-prior) & pollutant\_uni & $0.024$ & $8.460$ & $0.1$ \\ 
sparsePCE & pollutant\_uni & $0.010$ & $0.070$ & $0$ \\ 
BART & pollutant\_uni & $0.011$ & $6.592$ & $0$ \\ 
laGP & pollutant\_uni & $0.008$ & $22.280$ & $0$ \\[1.2ex]
KHAOS (ridge) & friedman20 & $0.938$ & $9.276$ & $0$ \\ 
KHAOS (g-prior) & friedman20 & $0.998$ & $9.855$ & $0$ \\ 
sparsePCE & friedman20 & $0.079$ & $12.534$ & $1$ \\ 
BART & friedman20 & $0.209$ & $6.556$ & $0$ \\ 
laGP & friedman20 & $1.354$ & $40.450$ & $0$ \\
\hline \\[-1.8ex] 
\end{tabular} 
\end{table} 

\begin{table}[!t] \centering 
  \caption{Results for the simulation study in the high noise $(NSR = 0.5)$ setting.} 
  \label{tab:tab3} 
\begin{tabular}{@{\extracolsep{5pt}} lcccc} 
\\[-1.8ex]\hline 
\hline \\[-1.8ex] 
Method & Function & Avg. CRPS & Avg. Time & Wtihin 1\% Rate \\ 
\hline \\[-1.8ex] 
KHAOS (ridge) & banana & $12.498$ & $7.032$ & $1$  \\ 
KHAOS (g-prior) & banana & $22.224$ & $12.967$ & $0$  \\ 
sparsePCE & banana & $76.245$ & $3.991$ & $0$  \\ 
BART & banana & $47.175$ & $6.972$ & $0$  \\ 
laGP & banana & $63.553$ & $20.782$ & $0$ \\[1.2ex]
KHAOS (ridge) & ishigami & $0.402$ & $8.118$ & $0.100$ \\ 
KHAOS (g-prior) & ishigami & $0.359$ & $12.362$ & $0.900$ \\ 
sparsePCE & ishigami & $87,626$ & $482.144$ & $0$ \\ 
BART & ishigami & $0.557$ & $7.425$ & $0$ \\ 
laGP & ishigami & $0.795$ & $21.441$  & $0$ \\[1.2ex] 
KHAOS (ridge) & rabbits & $0.035$ & $7.516$ & $0.100$  \\ 
KHAOS (g-prior) & rabbits & $0.033$ & $10.601$ &  $0.500$ \\ 
sparsePCE & rabbits & $2.659$ & $409.167$ & $0$  \\ 
BART & rabbits & $0.032$ & $7.254$ & $0.500$  \\ 
laGP & rabbits & $0.050$ & $23.226$  & $0$ \\[1.2ex] 
KHAOS (ridge) & pollutant\_uni & $0.084$ & $7.711$  & $0.300$ \\ 
KHAOS (g-prior) & pollutant\_uni & $0.071$ & $9.387$ & $0.800$  \\ 
sparsePCE & pollutant\_uni & $0.413$ & $635.362$ & $0$ \\ 
BART & pollutant\_uni & $0.094$ & $6.689$ & $0.100$ \\ 
laGP & pollutant\_uni & $0.254$ & $21.201$ & $0$ \\[1.2ex] 
KHAOS (ridge) & friedman20 & $1.059$ & $8.778$ & $0.200$  \\ 
KHAOS (g-prior) & friedman20 & $0.900$ & $9.112$ & $0.400$  \\ 
sparsePCE & friedman20 & $2.531$ & $294.623$ & $0$  \\ 
BART & friedman20 & $0.860$ & $7.368$ & $0.400$  \\ 
laGP & friedman20 & $2.317$ & $39.172$  & $0$ \\ 
\hline \\[-1.8ex] 
\end{tabular} 
\end{table}

For each test function, we generated a training set of $n = 1000$ points using maximin Latin hypercube sampling \citep{mckay1979}. Responses include additive noise under two settings: a noise-free emulation case ($NSR = 0$) and a high-noise regression case ($NSR = 0.5$). 

We evaluate each emulation method using continuous ranked probability scores (CRPS), a proper scoring rule that balances precision and accuracy of a distributional prediction \citep{gneiting2007strictly}. The CRPS is defined as
\begin{equation}
\text{CRPS}(F, y_\text{true}) = \int_{-\infty}^{\infty} \left( F(z) - \mathbf{1}\{z \geq y_\text{true}\} \right)^2 \, dz = \mathbb{E}_F |Y - y_\text{true}| - \frac{1}{2} \mathbb{E}_F |Y - Y^\prime|
\end{equation}
where $Y$ and $Y^\prime$ are independent draws from $F$. Each method is tested on an independent test set of size $1000$. All simulation scenarios are replicated 10 times with fresh designs and noise.

\subsection{Results}
A visual summary of the results for the noise free setting are given in \cref{fig:sub1}, which shows the average CRPS ranking of each emulator across the ten replications. Complete results including timing and raw CRPS averages are given in \cref{tab:tab2}. In the high-noise setting, equivalent figures and tables are given by \cref{fig:sub2} and \cref{tab:tab3}.

Some takeaways of this analysis include:
\begin{itemize}
    \item No single emulator is ever the best across all $5$ test functions.
    \item In the noise free setting, the KHAOS approach with a ridge prior has the best average CRPS rank. 
    \item In the high-noise setting, the KHAOS approach with a modified $g$-prior has the best average CRPS rank. This is likely due to the $g$-priors ability to reduce potential for overfitting.
    \item The Sparse PCE approach does reasonably well in the noise free setting (and always has the best CRPS for "friedman20") but appears to overfit in the high noise setting. 
    \item When $NSR = 0$, the laGP emulator performs well. When $NSR = 0.5$, BART demonstrates good performance. Both of these findings are consistent with previous work. 
\end{itemize}
While emulator performance is problem-dependent, KHAOS performs consistently well across functions and demonstrates robustness to both low- and high-noise settings. For additional figures, including boxplots of CRPS, heatmaps based on RMSE, and a Pareto plot comparing speed and accurayc, see Appendix E in the supplemental materials. 
\vfill

\section{Real Data Examples}
\label{sec:application}
\label{sec:realdata}
We illustrate the flexibility of KHAOS on two real datasets. The first is a physics-based computer model with $p = 6$ inputs, which simulates an exploding cylinder with a gold liner; see \cite{rumsey2025co} for details. The second is the UCI white wine quality dataset, where the response is ordinal \citep{cortez2009modeling}.

For the ordinal data, we follow the latent Gaussian approach described by \cite{hoff2009first}, applying KHAOS to the latent space to enable Sobol decompositions of variance. This implementation is available in the \texttt{ordinal\_khaos} function in the \texttt{khaos} package.  

\begin{figure}[ht]
  \centering
  \begin{subfigure}[t]{0.48\textwidth}
    \centering
    \includegraphics[width=\linewidth]{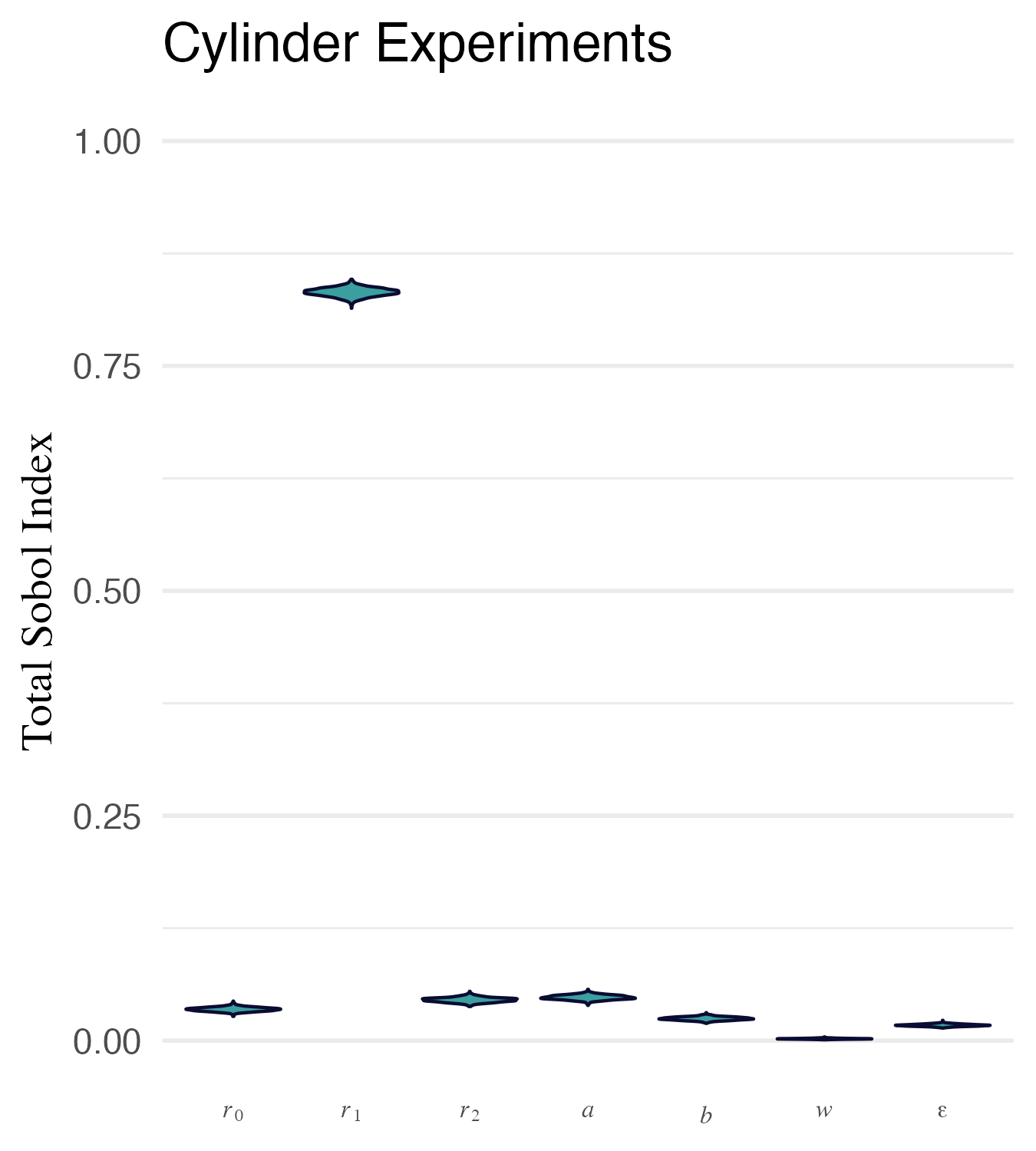}
    \caption{Total Sobol indices for the Cylinder Experiments dataset, showing high sensitivity to the input $r_1$. The variance not explainable by KHAOS (due to $\epsilon$) is negligible.}
    \label{fig:sub3}
  \end{subfigure}
  \hfill
  \begin{subfigure}[t]{0.48\textwidth}
    \centering
    \includegraphics[width=\linewidth]{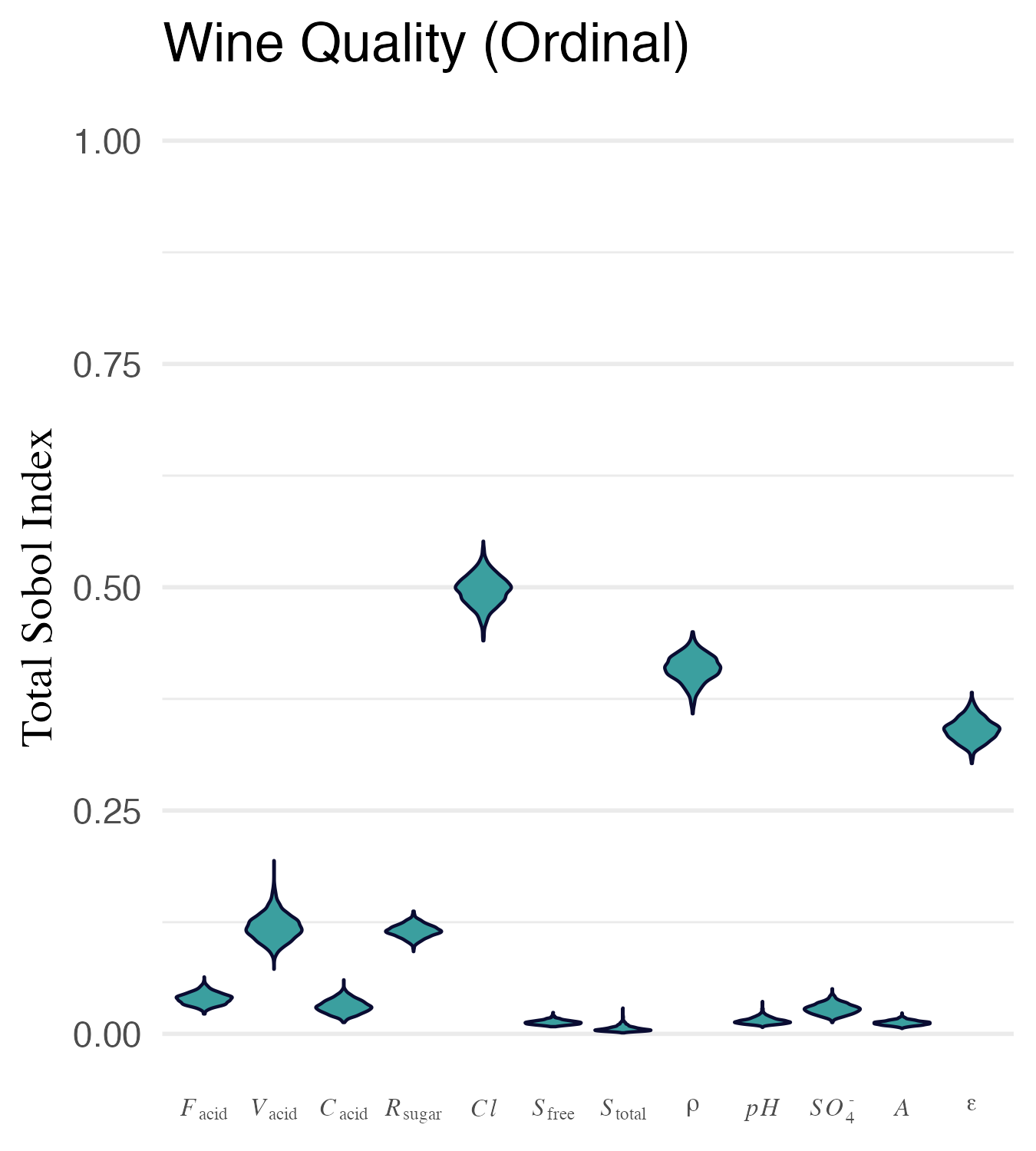}
    \caption{Total Sobol indices for the Wine Quality dataset with ordinal response. Several variables are deemed important, to varying degrees, and a substantial amount of the latent variance is left unexplained.}
    \label{fig:sub4}
  \end{subfigure}
  \caption{}
  \label{fig:combined2}
\end{figure}

\Cref{fig:sub3} shows the total Sobol indices for the Cylinder Experiments, with dominant sensitivity to input $r_1$ and negligible unexplained variance (denoted as $\epsilon$ in the each subpanel of \cref{fig:combined2}). In contrast, \cref{fig:sub4} shows that in the wine dataset, several inputs contribute meaningfully to the latent response, but a substantial portion of the variance remains unexplained.


\section{Conclusion}
\label{sec:conclusion}

There are many effective emulators available, and no single method works best across all problems. As suggested by no-free-lunch theorems, emulator performance depends on the structure of the function, noise levels, and the evaluation criteria. KHAOS is not a one-size-fits-all solution, but it is a robust and flexible tool that performs well across a range of settings.

Like other additive Bayesian methods (e.g., BASS, BPPR, BART), KHAOS models complex functions through structured basis expansions with full posterior inference. It builds on polynomial chaos ideas and naturally supports global sensitivity analysis via posterior Sobol indices (even in latent data settings). This leads to interpretable uncertainty quantification alongside competitive predictive accuracy. Future work might focus on extending the use of KHAOS for sensitivity studies via (e.g.) Shapley effects \citep{owen2014sobol} or dimension reduction via (e.g.) active subspaces \citep{constantine2015, rumsey2024discovering}. 

The \texttt{khaos} R package fills a gap in the R ecosystem by providing a fully Bayesian PCE implementation with support for uncertainty quantification and sensitivity analysis—tools that are useful in both emulator evaluation and scientific applications.


\bibliography{WileyNJD-AMA}

\begin{thebibliography}{}

\bibitem [\protect \citeauthoryear {%
Bliznyuk%
\ \protect \BOthers {.}}{%
Bliznyuk%
\ \protect \BOthers {.}}{%
{\protect \APACyear {2008}}%
}]{%
bliznyuk2008bayesian}
\APACinsertmetastar {%
bliznyuk2008bayesian}%
\begin{APACrefauthors}%
Bliznyuk, N.%
, Ruppert, D.%
, Shoemaker, C.%
, Regis, R.%
, Wild, S.%
\BCBL {}\ \BBA {} Mugunthan, P.%
\end{APACrefauthors}%
\unskip\
\newblock
\APACrefYearMonthDay{2008}{}{}.
\newblock
{\BBOQ}\APACrefatitle {Bayesian calibration and uncertainty analysis for
  computationally expensive models using optimization and radial basis function
  approximation} {Bayesian calibration and uncertainty analysis for
  computationally expensive models using optimization and radial basis function
  approximation}.{\BBCQ}
\newblock
\APACjournalVolNumPages{Journal of Computational and Graphical
  Statistics}{17}{2}{270--294}.
\PrintBackRefs{\CurrentBib}

\bibitem [\protect \citeauthoryear {%
Chipman%
, George%
\BCBL {}\ \BBA {} McCulloch%
}{%
Chipman%
\ \protect \BOthers {.}}{%
{\protect \APACyear {2010}}%
}]{%
chipman2010bart}
\APACinsertmetastar {%
chipman2010bart}%
\begin{APACrefauthors}%
Chipman, H\BPBI A.%
, George, E\BPBI I.%
\BCBL {}\ \BBA {} McCulloch, R\BPBI E.%
\end{APACrefauthors}%
\unskip\
\newblock
\APACrefYearMonthDay{2010}{}{}.
\newblock
{\BBOQ}\APACrefatitle {BART: Bayesian additive regression trees} {Bart:
  Bayesian additive regression trees}.{\BBCQ}
\newblock
\APACjournalVolNumPages{}{4}{1}{266}.
\PrintBackRefs{\CurrentBib}

\bibitem [\protect \citeauthoryear {%
Collins%
, Francom%
\BCBL {}\ \BBA {} Rumsey%
}{%
Collins%
\ \protect \BOthers {.}}{%
{\protect \APACyear {2024}}%
}]{%
collins2024bayesian}
\APACinsertmetastar {%
collins2024bayesian}%
\begin{APACrefauthors}%
Collins, G.%
, Francom, D.%
\BCBL {}\ \BBA {} Rumsey, K.%
\end{APACrefauthors}%
\unskip\
\newblock
\APACrefYearMonthDay{2024}{}{}.
\newblock
{\BBOQ}\APACrefatitle {Bayesian projection pursuit regression} {Bayesian
  projection pursuit regression}.{\BBCQ}
\newblock
\APACjournalVolNumPages{Statistics and Computing}{34}{1}{29}.
\PrintBackRefs{\CurrentBib}

\bibitem [\protect \citeauthoryear {%
Constantine%
}{%
Constantine%
}{%
{\protect \APACyear {2015}}%
}]{%
constantine2015}
\APACinsertmetastar {%
constantine2015}%
\begin{APACrefauthors}%
Constantine, P\BPBI G.%
\end{APACrefauthors}%
\unskip\
\newblock
\APACrefYear{2015}.
\newblock
\APACrefbtitle {Active subspaces: Emerging ideas for dimension reduction in
  parameter studies} {Active subspaces: Emerging ideas for dimension reduction
  in parameter studies}\ (\BVOL~2).
\newblock
\APACaddressPublisher{}{SIAM}.
\PrintBackRefs{\CurrentBib}

\bibitem [\protect \citeauthoryear {%
Cortez%
, Cerdeira%
, Almeida%
, Matos%
\BCBL {}\ \BBA {} Reis%
}{%
Cortez%
\ \protect \BOthers {.}}{%
{\protect \APACyear {2009}}%
}]{%
cortez2009modeling}
\APACinsertmetastar {%
cortez2009modeling}%
\begin{APACrefauthors}%
Cortez, P.%
, Cerdeira, A.%
, Almeida, F.%
, Matos, T.%
\BCBL {}\ \BBA {} Reis, J.%
\end{APACrefauthors}%
\unskip\
\newblock
\APACrefYearMonthDay{2009}{}{}.
\newblock
{\BBOQ}\APACrefatitle {Modeling wine preferences by data mining from
  physicochemical properties} {Modeling wine preferences by data mining from
  physicochemical properties}.{\BBCQ}
\newblock
\APACjournalVolNumPages{Decision support systems}{47}{4}{547--553}.
\PrintBackRefs{\CurrentBib}

\bibitem [\protect \citeauthoryear {%
Denison%
, Mallick%
\BCBL {}\ \BBA {} Smith%
}{%
Denison%
\ \protect \BOthers {.}}{%
{\protect \APACyear {1998}}%
}]{%
denison1998bayesian}
\APACinsertmetastar {%
denison1998bayesian}%
\begin{APACrefauthors}%
Denison, D\BPBI G.%
, Mallick, B\BPBI K.%
\BCBL {}\ \BBA {} Smith, A\BPBI F.%
\end{APACrefauthors}%
\unskip\
\newblock
\APACrefYearMonthDay{1998}{}{}.
\newblock
{\BBOQ}\APACrefatitle {Bayesian mars} {Bayesian mars}.{\BBCQ}
\newblock
\APACjournalVolNumPages{Statistics and Computing}{8}{}{337--346}.
\PrintBackRefs{\CurrentBib}

\bibitem [\protect \citeauthoryear {%
Francom%
\ \BBA {} Sans{\'o}%
}{%
Francom%
\ \BBA {} Sans{\'o}%
}{%
{\protect \APACyear {2020}}%
}]{%
francom2020bass}
\APACinsertmetastar {%
francom2020bass}%
\begin{APACrefauthors}%
Francom, D.%
\BCBT {}\ \BBA {} Sans{\'o}, B.%
\end{APACrefauthors}%
\unskip\
\newblock
\APACrefYearMonthDay{2020}{}{}.
\newblock
{\BBOQ}\APACrefatitle {BASS: An R package for fitting and performing
  sensitivity analysis of Bayesian adaptive spline surfaces} {Bass: An r
  package for fitting and performing sensitivity analysis of bayesian adaptive
  spline surfaces}.{\BBCQ}
\newblock
\APACjournalVolNumPages{Journal of Statistical Software}{94}{LA-UR-20-23587}{}.
\PrintBackRefs{\CurrentBib}

\bibitem [\protect \citeauthoryear {%
Francom%
, Sans{\'o}%
, Kupresanin%
\BCBL {}\ \BBA {} Johannesson%
}{%
Francom%
\ \protect \BOthers {.}}{%
{\protect \APACyear {2018}}%
}]{%
francom2018}
\APACinsertmetastar {%
francom2018}%
\begin{APACrefauthors}%
Francom, D.%
, Sans{\'o}, B.%
, Kupresanin, A.%
\BCBL {}\ \BBA {} Johannesson, G.%
\end{APACrefauthors}%
\unskip\
\newblock
\APACrefYearMonthDay{2018}{}{}.
\newblock
{\BBOQ}\APACrefatitle {Sensitivity analysis and emulation for functional data
  using Bayesian adaptive splines} {Sensitivity analysis and emulation for
  functional data using bayesian adaptive splines}.{\BBCQ}
\newblock
\APACjournalVolNumPages{Statistica Sinica}{}{}{791--816}.
\PrintBackRefs{\CurrentBib}

\bibitem [\protect \citeauthoryear {%
Friedman%
}{%
Friedman%
}{%
{\protect \APACyear {1991}}%
}]{%
friedman1991}
\APACinsertmetastar {%
friedman1991}%
\begin{APACrefauthors}%
Friedman, J\BPBI H.%
\end{APACrefauthors}%
\unskip\
\newblock
\APACrefYearMonthDay{1991}{}{}.
\newblock
{\BBOQ}\APACrefatitle {Multivariate adaptive regression splines} {Multivariate
  adaptive regression splines}.{\BBCQ}
\newblock
\APACjournalVolNumPages{The annals of statistics}{}{}{1--67}.
\PrintBackRefs{\CurrentBib}

\bibitem [\protect \citeauthoryear {%
Ghanem%
, Spanos%
, Ghanem%
\BCBL {}\ \BBA {} Spanos%
}{%
Ghanem%
\ \protect \BOthers {.}}{%
{\protect \APACyear {1991}}%
}]{%
ghanem1991stochastic}
\APACinsertmetastar {%
ghanem1991stochastic}%
\begin{APACrefauthors}%
Ghanem, R\BPBI G.%
, Spanos, P\BPBI D.%
, Ghanem, R\BPBI G.%
\BCBL {}\ \BBA {} Spanos, P\BPBI D.%
\end{APACrefauthors}%
\unskip\
\newblock
\APACrefYearMonthDay{1991}{}{}.
\newblock
{\BBOQ}\APACrefatitle {Stochastic finite element method: Response statistics}
  {Stochastic finite element method: Response statistics}.{\BBCQ}
\newblock
\APACjournalVolNumPages{Stochastic finite elements: a spectral
  approach}{}{}{101--119}.
\PrintBackRefs{\CurrentBib}

\bibitem [\protect \citeauthoryear {%
Gneiting%
\ \BBA {} Raftery%
}{%
Gneiting%
\ \BBA {} Raftery%
}{%
{\protect \APACyear {2007}}%
}]{%
gneiting2007strictly}
\APACinsertmetastar {%
gneiting2007strictly}%
\begin{APACrefauthors}%
Gneiting, T.%
\BCBT {}\ \BBA {} Raftery, A\BPBI E.%
\end{APACrefauthors}%
\unskip\
\newblock
\APACrefYearMonthDay{2007}{}{}.
\newblock
{\BBOQ}\APACrefatitle {Strictly proper scoring rules, prediction, and
  estimation} {Strictly proper scoring rules, prediction, and
  estimation}.{\BBCQ}
\newblock
\APACjournalVolNumPages{Journal of the American statistical
  Association}{102}{477}{359--378}.
\PrintBackRefs{\CurrentBib}

\bibitem [\protect \citeauthoryear {%
Gotelli%
, Ellison%
\BCBL {}\ \protect \BOthers {.}}{%
Gotelli%
\ \protect \BOthers {.}}{%
{\protect \APACyear {2004}}%
}]{%
gotelli2004primer}
\APACinsertmetastar {%
gotelli2004primer}%
\begin{APACrefauthors}%
Gotelli, N\BPBI J.%
, Ellison, A\BPBI M.%
\BCBL {}\ \BOthersPeriod {.}\end{APACrefauthors}%
\unskip\
\newblock
\APACrefYear{2004}.
\newblock
\APACrefbtitle {A primer of ecological statistics} {A primer of ecological
  statistics}\ (\BVOL~1).
\newblock
\APACaddressPublisher{}{Sinauer Associates Sunderland}.
\PrintBackRefs{\CurrentBib}

\bibitem [\protect \citeauthoryear {%
Gramacy%
\ \BBA {} Apley%
}{%
Gramacy%
\ \BBA {} Apley%
}{%
{\protect \APACyear {2015}}%
}]{%
gramacy2015}
\APACinsertmetastar {%
gramacy2015}%
\begin{APACrefauthors}%
Gramacy, R\BPBI B.%
\BCBT {}\ \BBA {} Apley, D\BPBI W.%
\end{APACrefauthors}%
\unskip\
\newblock
\APACrefYearMonthDay{2015}{}{}.
\newblock
{\BBOQ}\APACrefatitle {Local Gaussian process approximation for large computer
  experiments} {Local gaussian process approximation for large computer
  experiments}.{\BBCQ}
\newblock
\APACjournalVolNumPages{Journal of Computational and Graphical
  Statistics}{24}{2}{561--578}.
\PrintBackRefs{\CurrentBib}

\bibitem [\protect \citeauthoryear {%
Green%
\ \BBA {} Mira%
}{%
Green%
\ \BBA {} Mira%
}{%
{\protect \APACyear {2001}}%
}]{%
green2001delayed}
\APACinsertmetastar {%
green2001delayed}%
\begin{APACrefauthors}%
Green, P\BPBI J.%
\BCBT {}\ \BBA {} Mira, A.%
\end{APACrefauthors}%
\unskip\
\newblock
\APACrefYearMonthDay{2001}{}{}.
\newblock
{\BBOQ}\APACrefatitle {Delayed rejection in reversible jump
  Metropolis--Hastings} {Delayed rejection in reversible jump
  metropolis--hastings}.{\BBCQ}
\newblock
\APACjournalVolNumPages{Biometrika}{88}{4}{1035--1053}.
\PrintBackRefs{\CurrentBib}

\bibitem [\protect \citeauthoryear {%
Hoff%
}{%
Hoff%
}{%
{\protect \APACyear {2009}}%
}]{%
hoff2009first}
\APACinsertmetastar {%
hoff2009first}%
\begin{APACrefauthors}%
Hoff, P\BPBI D.%
\end{APACrefauthors}%
\unskip\
\newblock
\APACrefYear{2009}.
\newblock
\APACrefbtitle {A first course in Bayesian statistical methods} {A first course
  in bayesian statistical methods}\ (\BVOL~580).
\newblock
\APACaddressPublisher{}{Springer}.
\PrintBackRefs{\CurrentBib}

\bibitem [\protect \citeauthoryear {%
Ishigami%
\ \BBA {} Homma%
}{%
Ishigami%
\ \BBA {} Homma%
}{%
{\protect \APACyear {1990}}%
}]{%
ishigami1990importance}
\APACinsertmetastar {%
ishigami1990importance}%
\begin{APACrefauthors}%
Ishigami, T.%
\BCBT {}\ \BBA {} Homma, T.%
\end{APACrefauthors}%
\unskip\
\newblock
\APACrefYearMonthDay{1990}{}{}.
\newblock
{\BBOQ}\APACrefatitle {An importance quantification technique in uncertainty
  analysis for computer models} {An importance quantification technique in
  uncertainty analysis for computer models}.{\BBCQ}
\newblock
\BIn{} \APACrefbtitle {[1990] Proceedings. First international symposium on
  uncertainty modeling and analysis} {[1990] proceedings. first international
  symposium on uncertainty modeling and analysis}\ (\BPGS\ 398--403).
\PrintBackRefs{\CurrentBib}

\bibitem [\protect \citeauthoryear {%
Liang%
, Paulo%
, Molina%
, Clyde%
\BCBL {}\ \BBA {} Berger%
}{%
Liang%
\ \protect \BOthers {.}}{%
{\protect \APACyear {2008}}%
}]{%
liang2008mixtures}
\APACinsertmetastar {%
liang2008mixtures}%
\begin{APACrefauthors}%
Liang, F.%
, Paulo, R.%
, Molina, G.%
, Clyde, M\BPBI A.%
\BCBL {}\ \BBA {} Berger, J\BPBI O.%
\end{APACrefauthors}%
\unskip\
\newblock
\APACrefYearMonthDay{2008}{}{}.
\newblock
{\BBOQ}\APACrefatitle {Mixtures of g priors for Bayesian variable selection}
  {Mixtures of g priors for bayesian variable selection}.{\BBCQ}
\newblock
\APACjournalVolNumPages{Journal of the American Statistical
  Association}{103}{481}{410--423}.
\PrintBackRefs{\CurrentBib}

\bibitem [\protect \citeauthoryear {%
Lüthen%
, Marelli%
\BCBL {}\ \BBA {} Sudret%
}{%
Lüthen%
\ \protect \BOthers {.}}{%
{\protect \APACyear {2021}}%
}]{%
luthen2021sparse}
\APACinsertmetastar {%
luthen2021sparse}%
\begin{APACrefauthors}%
Lüthen, N.%
, Marelli, S.%
\BCBL {}\ \BBA {} Sudret, B.%
\end{APACrefauthors}%
\unskip\
\newblock
\APACrefYearMonthDay{2021}{}{}.
\newblock
{\BBOQ}\APACrefatitle {Sparse polynomial chaos expansions: Literature survey
  and benchmark} {Sparse polynomial chaos expansions: Literature survey and
  benchmark}.{\BBCQ}
\newblock
\APACjournalVolNumPages{SIAM/ASA Journal on Uncertainty
  Quantification}{9}{2}{593--649}.
\PrintBackRefs{\CurrentBib}

\bibitem [\protect \citeauthoryear {%
McKay%
, Beckman%
\BCBL {}\ \BBA {} Conover%
}{%
McKay%
\ \protect \BOthers {.}}{%
{\protect \APACyear {1979}}%
}]{%
mckay1979}
\APACinsertmetastar {%
mckay1979}%
\begin{APACrefauthors}%
McKay, M\BPBI D.%
, Beckman, R\BPBI J.%
\BCBL {}\ \BBA {} Conover, W\BPBI J.%
\end{APACrefauthors}%
\unskip\
\newblock
\APACrefYearMonthDay{1979}{}{}.
\newblock
{\BBOQ}\APACrefatitle {Comparison of three methods for selecting values of
  input variables in the analysis of output from a computer code} {Comparison
  of three methods for selecting values of input variables in the analysis of
  output from a computer code}.{\BBCQ}
\newblock
\APACjournalVolNumPages{Technometrics}{21}{2}{239--245}.
\PrintBackRefs{\CurrentBib}

\bibitem [\protect \citeauthoryear {%
{Nature Methods Editorial}%
}{%
{Nature Methods Editorial}%
}{%
{\protect \APACyear {2011}}%
}]{%
nuap2011}
\APACinsertmetastar {%
nuap2011}%
\begin{APACrefauthors}%
{Nature Methods Editorial}.%
\end{APACrefauthors}%
\unskip\
\newblock
\APACrefYearMonthDay{2011}{}{}.
\newblock
{\BBOQ}\APACrefatitle {{NUAP} (no unnecessary acronyms please)} {{NUAP} (no
  unnecessary acronyms please)}.{\BBCQ}
\newblock
\APACjournalVolNumPages{Nature Methods}{8}{}{521}.
\newblock
\begin{APACrefDOI} 10.1038/nmeth.1646 \end{APACrefDOI}
\PrintBackRefs{\CurrentBib}

\bibitem [\protect \citeauthoryear {%
Nott%
, Kuk%
\BCBL {}\ \BBA {} Duc%
}{%
Nott%
\ \protect \BOthers {.}}{%
{\protect \APACyear {2005}}%
}]{%
nott2005efficient}
\APACinsertmetastar {%
nott2005efficient}%
\begin{APACrefauthors}%
Nott, D\BPBI J.%
, Kuk, A\BPBI Y.%
\BCBL {}\ \BBA {} Duc, H.%
\end{APACrefauthors}%
\unskip\
\newblock
\APACrefYearMonthDay{2005}{}{}.
\newblock
{\BBOQ}\APACrefatitle {Efficient sampling schemes for Bayesian MARS models with
  many predictors} {Efficient sampling schemes for bayesian mars models with
  many predictors}.{\BBCQ}
\newblock
\APACjournalVolNumPages{Statistics and Computing}{15}{}{93--101}.
\PrintBackRefs{\CurrentBib}

\bibitem [\protect \citeauthoryear {%
Novak%
\ \BBA {} Novak%
}{%
Novak%
\ \BBA {} Novak%
}{%
{\protect \APACyear {2018}}%
}]{%
novak2018polynomial}
\APACinsertmetastar {%
novak2018polynomial}%
\begin{APACrefauthors}%
Novak, L.%
\BCBT {}\ \BBA {} Novak, D.%
\end{APACrefauthors}%
\unskip\
\newblock
\APACrefYearMonthDay{2018}{}{}.
\newblock
{\BBOQ}\APACrefatitle {Polynomial chaos expansion for surrogate modelling:
  Theory and software} {Polynomial chaos expansion for surrogate modelling:
  Theory and software}.{\BBCQ}
\newblock
\APACjournalVolNumPages{Beton-und Stahlbetonbau}{113}{}{27--32}.
\PrintBackRefs{\CurrentBib}

\bibitem [\protect \citeauthoryear {%
Owen%
}{%
Owen%
}{%
{\protect \APACyear {2014}}%
}]{%
owen2014sobol}
\APACinsertmetastar {%
owen2014sobol}%
\begin{APACrefauthors}%
Owen, A\BPBI B.%
\end{APACrefauthors}%
\unskip\
\newblock
\APACrefYearMonthDay{2014}{}{}.
\newblock
{\BBOQ}\APACrefatitle {Sobol'indices and Shapley value} {Sobol'indices and
  shapley value}.{\BBCQ}
\newblock
\APACjournalVolNumPages{SIAM/ASA Journal on Uncertainty
  Quantification}{2}{1}{245--251}.
\PrintBackRefs{\CurrentBib}

\bibitem [\protect \citeauthoryear {%
O’Hagan%
\ \protect \BOthers {.}}{%
O’Hagan%
\ \protect \BOthers {.}}{%
{\protect \APACyear {2013}}%
}]{%
o2013polynomial}
\APACinsertmetastar {%
o2013polynomial}%
\begin{APACrefauthors}%
O’Hagan, A.%
\BCBT {}\ \BOthersPeriod {.}
\end{APACrefauthors}%
\unskip\
\newblock
\APACrefYearMonthDay{2013}{}{}.
\newblock
{\BBOQ}\APACrefatitle {Polynomial chaos: A tutorial and critique from a
  statistician’s perspective} {Polynomial chaos: A tutorial and critique from
  a statistician’s perspective}.{\BBCQ}
\newblock
\APACjournalVolNumPages{SIAM/ASA J. Uncertainty Quantification}{20}{}{1--20}.
\PrintBackRefs{\CurrentBib}

\bibitem [\protect \citeauthoryear {%
Robert%
\ \protect \BOthers {.}}{%
Robert%
\ \protect \BOthers {.}}{%
{\protect \APACyear {2007}}%
}]{%
robert2007bayesian}
\APACinsertmetastar {%
robert2007bayesian}%
\begin{APACrefauthors}%
Robert, C\BPBI P.%
\BCBT {}\ \BOthersPeriod {.}
\end{APACrefauthors}%
\unskip\
\newblock
\APACrefYear{2007}.
\newblock
\APACrefbtitle {The Bayesian choice: from decision-theoretic foundations to
  computational implementation} {The bayesian choice: from decision-theoretic
  foundations to computational implementation}\ (\BVOL~2).
\newblock
\APACaddressPublisher{}{Springer}.
\PrintBackRefs{\CurrentBib}

\bibitem [\protect \citeauthoryear {%
K.~Rumsey%
}{%
K.~Rumsey%
}{%
{\protect \APACyear {2023}}%
}]{%
rumsey2023duqling}
\APACinsertmetastar {%
rumsey2023duqling}%
\begin{APACrefauthors}%
Rumsey, K.%
\end{APACrefauthors}%
\unskip\
\newblock
\APACrefYearMonthDay{2023}{}{}.
\newblock
\APACrefbtitle {duqling} {duqling}\ \APACbVolEdTR{}{\BTR{}}.
\newblock
\APACaddressInstitution{}{Los Alamos National Laboratory (LANL), Los Alamos, NM
  (United States)}.
\PrintBackRefs{\CurrentBib}

\bibitem [\protect \citeauthoryear {%
K.~Rumsey%
, Francom%
\BCBL {}\ \BBA {} Vander~Wiel%
}{%
K.~Rumsey%
\ \protect \BOthers {.}}{%
{\protect \APACyear {2024}}%
}]{%
rumsey2024discovering}
\APACinsertmetastar {%
rumsey2024discovering}%
\begin{APACrefauthors}%
Rumsey, K.%
, Francom, D.%
\BCBL {}\ \BBA {} Vander~Wiel, S.%
\end{APACrefauthors}%
\unskip\
\newblock
\APACrefYearMonthDay{2024}{}{}.
\newblock
{\BBOQ}\APACrefatitle {Discovering active subspaces for high-dimensional
  computer models} {Discovering active subspaces for high-dimensional computer
  models}.{\BBCQ}
\newblock
\APACjournalVolNumPages{Journal of Computational and Graphical
  Statistics}{33}{3}{896--908}.
\PrintBackRefs{\CurrentBib}

\bibitem [\protect \citeauthoryear {%
K\BPBI N.~Rumsey%
, Francom%
\BCBL {}\ \BBA {} Shen%
}{%
K\BPBI N.~Rumsey%
\ \protect \BOthers {.}}{%
{\protect \APACyear {2024}}%
}]{%
rumsey2024generalized}
\APACinsertmetastar {%
rumsey2024generalized}%
\begin{APACrefauthors}%
Rumsey, K\BPBI N.%
, Francom, D.%
\BCBL {}\ \BBA {} Shen, A.%
\end{APACrefauthors}%
\unskip\
\newblock
\APACrefYearMonthDay{2024}{}{}.
\newblock
{\BBOQ}\APACrefatitle {Generalized Bayesian MARS: Tools for Stochastic Computer
  Model Emulation} {Generalized bayesian mars: Tools for stochastic computer
  model emulation}.{\BBCQ}
\newblock
\APACjournalVolNumPages{SIAM/ASA Journal on Uncertainty
  Quantification}{12}{2}{646--666}.
\PrintBackRefs{\CurrentBib}

\bibitem [\protect \citeauthoryear {%
K\BPBI N.~Rumsey%
, Hardy%
, Ahrens%
\BCBL {}\ \BBA {} Vander~Wiel%
}{%
K\BPBI N.~Rumsey%
\ \protect \BOthers {.}}{%
{\protect \APACyear {2025}}%
}]{%
rumsey2025co}
\APACinsertmetastar {%
rumsey2025co}%
\begin{APACrefauthors}%
Rumsey, K\BPBI N.%
, Hardy, Z\BPBI K.%
, Ahrens, C.%
\BCBL {}\ \BBA {} Vander~Wiel, S.%
\end{APACrefauthors}%
\unskip\
\newblock
\APACrefYearMonthDay{2025}{}{}.
\newblock
{\BBOQ}\APACrefatitle {Co-Active Subspace Methods for the Joint Analysis of
  Adjacent Computer Models} {Co-active subspace methods for the joint analysis
  of adjacent computer models}.{\BBCQ}
\newblock
\APACjournalVolNumPages{Technometrics}{67}{1}{133--146}.
\PrintBackRefs{\CurrentBib}

\bibitem [\protect \citeauthoryear {%
Shao%
, Younes%
, Fahs%
\BCBL {}\ \BBA {} Mara%
}{%
Shao%
\ \protect \BOthers {.}}{%
{\protect \APACyear {2017}}%
}]{%
shao2017bayesian}
\APACinsertmetastar {%
shao2017bayesian}%
\begin{APACrefauthors}%
Shao, Q.%
, Younes, A.%
, Fahs, M.%
\BCBL {}\ \BBA {} Mara, T\BPBI A.%
\end{APACrefauthors}%
\unskip\
\newblock
\APACrefYearMonthDay{2017}{}{}.
\newblock
{\BBOQ}\APACrefatitle {Bayesian sparse polynomial chaos expansion for global
  sensitivity analysis} {Bayesian sparse polynomial chaos expansion for global
  sensitivity analysis}.{\BBCQ}
\newblock
\APACjournalVolNumPages{Computer Methods in Applied Mechanics and
  Engineering}{318}{}{474--496}.
\PrintBackRefs{\CurrentBib}

\bibitem [\protect \citeauthoryear {%
Sobol%
}{%
Sobol%
}{%
{\protect \APACyear {2001}}%
}]{%
sobol2001global}
\APACinsertmetastar {%
sobol2001global}%
\begin{APACrefauthors}%
Sobol, I\BPBI M.%
\end{APACrefauthors}%
\unskip\
\newblock
\APACrefYearMonthDay{2001}{}{}.
\newblock
{\BBOQ}\APACrefatitle {Global sensitivity indices for nonlinear mathematical
  models and their Monte Carlo estimates} {Global sensitivity indices for
  nonlinear mathematical models and their monte carlo estimates}.{\BBCQ}
\newblock
\APACjournalVolNumPages{Mathematics and computers in
  simulation}{55}{1-3}{271--280}.
\PrintBackRefs{\CurrentBib}

\bibitem [\protect \citeauthoryear {%
Sudret%
}{%
Sudret%
}{%
{\protect \APACyear {2008}}%
}]{%
sudret2008global}
\APACinsertmetastar {%
sudret2008global}%
\begin{APACrefauthors}%
Sudret, B.%
\end{APACrefauthors}%
\unskip\
\newblock
\APACrefYearMonthDay{2008}{}{}.
\newblock
{\BBOQ}\APACrefatitle {Global sensitivity analysis using polynomial chaos
  expansions} {Global sensitivity analysis using polynomial chaos
  expansions}.{\BBCQ}
\newblock
\APACjournalVolNumPages{Reliability engineering \& system
  safety}{93}{7}{964--979}.
\PrintBackRefs{\CurrentBib}

\bibitem [\protect \citeauthoryear {%
Wiener%
}{%
Wiener%
}{%
{\protect \APACyear {1938}}%
}]{%
wiener1938homogeneous}
\APACinsertmetastar {%
wiener1938homogeneous}%
\begin{APACrefauthors}%
Wiener, N.%
\end{APACrefauthors}%
\unskip\
\newblock
\APACrefYearMonthDay{1938}{}{}.
\newblock
{\BBOQ}\APACrefatitle {The homogeneous chaos} {The homogeneous chaos}.{\BBCQ}
\newblock
\APACjournalVolNumPages{American Journal of Mathematics}{60}{4}{897--936}.
\PrintBackRefs{\CurrentBib}

\bibitem [\protect \citeauthoryear {%
Xiu%
\ \BBA {} Karniadakis%
}{%
Xiu%
\ \BBA {} Karniadakis%
}{%
{\protect \APACyear {2002}}%
}]{%
xiu2002wiener}
\APACinsertmetastar {%
xiu2002wiener}%
\begin{APACrefauthors}%
Xiu, D.%
\BCBT {}\ \BBA {} Karniadakis, G\BPBI E.%
\end{APACrefauthors}%
\unskip\
\newblock
\APACrefYearMonthDay{2002}{}{}.
\newblock
{\BBOQ}\APACrefatitle {The Wiener--Askey polynomial chaos for stochastic
  differential equations} {The wiener--askey polynomial chaos for stochastic
  differential equations}.{\BBCQ}
\newblock
\APACjournalVolNumPages{SIAM journal on scientific computing}{24}{2}{619--644}.
\PrintBackRefs{\CurrentBib}

\bibitem [\protect \citeauthoryear {%
Zellner%
}{%
Zellner%
}{%
{\protect \APACyear {1986}}%
}]{%
zellner1986assessing}
\APACinsertmetastar {%
zellner1986assessing}%
\begin{APACrefauthors}%
Zellner, A.%
\end{APACrefauthors}%
\unskip\
\newblock
\APACrefYearMonthDay{1986}{}{}.
\newblock
{\BBOQ}\APACrefatitle {On assessing prior distributions and Bayesian regression
  analysis with g-prior distributions} {On assessing prior distributions and
  bayesian regression analysis with g-prior distributions}.{\BBCQ}
\newblock
\APACjournalVolNumPages{Bayesian inference and decision techniques}{}{}{}.
\PrintBackRefs{\CurrentBib}

\bibitem [\protect \citeauthoryear {%
Zhang%
, Huang%
, Gan%
, Karmaus%
\BCBL {}\ \BBA {} Sabo-Attwood%
}{%
Zhang%
\ \protect \BOthers {.}}{%
{\protect \APACyear {2016}}%
}]{%
zhang2016two}
\APACinsertmetastar {%
zhang2016two}%
\begin{APACrefauthors}%
Zhang, H.%
, Huang, X.%
, Gan, J.%
, Karmaus, W.%
\BCBL {}\ \BBA {} Sabo-Attwood, T.%
\end{APACrefauthors}%
\unskip\
\newblock
\APACrefYearMonthDay{2016}{}{}.
\newblock
{\BBOQ}\APACrefatitle {A two-component g-prior for variable selection} {A
  two-component g-prior for variable selection}.{\BBCQ}
\newblock

\PrintBackRefs{\CurrentBib}

\end{thebibliography}

\section*{Supporting Information}
The supporting information for this manuscript includes the \texttt{khaos} R package which is hosted at \url{https://github.com/knrumsey/khaos}, code to recreate all figures in this manuscript (hosted at \url{https://github.com/knrumsey/duqling_results}), and the document \texttt{SM\_khaos.pdf} with sections:
\begin{itemize}
    \item {\bf Appendix A. Enrichment Strategies}: Gives suggestions for alternate enrichment strategies in sparse PCE which are available in the \texttt{khaos} package. 
    \item {\bf Appendix B. Marginal Likelihood and Model Selection}: Additional information about the modified $g$-prior and a discussion on how it could be used in the sparse PCE agorithm of \cite{shao2017bayesian}.
    \item {\bf Appendix C. The Coinflip Proposal}: Additional details for the coinflip proposal discussed in \cref{sec:adaptive_proposal}. 
    \item {\bf Appendix D. Details of the Laplace Approximation}: Mathematical details surrounding the Laplace approximation to the conditional posterior of $g_0^2$. 
    \item {\bf Appendix E. Simulation Study: Additional Analysis}: Additional plots for the simulation study of \cref{sec:simulations}, not shown here for brevity.
\end{itemize}

\section*{Acknowledgments}

The authors thank Dr. Thierry Mara for his helpful discussions and correspondence during the development of this work.
\end{document}